\documentclass{ieeetj}
\usepackage{cite}
\usepackage{amsmath,amssymb,amsfonts}
\usepackage{algorithmic}
\usepackage{graphicx,color}
\usepackage{textcomp}
\usepackage{xcolor}
\usepackage{hyperref}
\hypersetup{hidelinks}
\usepackage{algorithm,algorithmic}
\def\BibTeX{{\rm B\kern-.05em{\sc i\kern-.025em b}\kern-.08em
    T\kern-.1667em\lower.7ex\hbox{E}\kern-.125emX}}
\AtBeginDocument{\definecolor{tmlcncolor}{cmyk}{0.93,0.59,0.15,0.02}\definecolor{NavyBlue}{RGB}{0,86,125}}
\usepackage{multirow}
\usepackage{amsmath,amssymb,amsfonts}
\usepackage{graphicx}
\usepackage{textcomp}
\usepackage{multirow}
\usepackage{booktabs}
\usepackage{capt-of}
\usepackage[table]{xcolor}
\usepackage{booktabs}
\usepackage{multirow}
\usepackage{capt-of}
\usepackage{stfloats}
\usepackage{hyperref}
\usepackage{pifont}
\usepackage{tcolorbox}
\usepackage{amsmath,graphicx,hyperref,orcidlink}

\def\OJlogo{}
\def\seclogo{}

\begin{document}
\raggedbottom
\bstctlcite{IEEEexample:BSTcontrol}

\markboth{}{Author {et al.}}

\title{TRACE-EVC: Text-Guided Relative Affective Control for Zero-Shot Emotional Voice Conversion}
\author{Zihan Zhang\orcidlink{0009-0004-1168-2163}, Shreeram Suresh Chandra\orcidlink{0000-0002-1085-2544}, Zongyang Du\orcidlink{0009-0006-5833-0071}, Xiutian Zhao\orcidlink{}, Aurosweta Mahapatra\orcidlink{0009-0001-3057-2021}, Hao Zhang\orcidlink{0009-0009-6205-2520}, Philipp Koehn\orcidlink{0000-0003-1565-064X} and Berrak Sisman\orcidlink{0000-0001-8078-3305}}
\affil{Center for Language and Speech Processing (CLSP), Johns Hopkins University, USA}
\authornote{This work is supported by the National Science Foundation (NSF) CAREER Award IIS-2533652 and by the National Research Foundation, Singapore, and the Ministry of Digital Development and Information under the Singapore Global AI Visiting Professorship (AIVP-2026-010). Any opinions, findings, conclusions, or recommendations expressed in this material are those of the authors and do not necessarily reflect the views of the National Research Foundation, Singapore, or the Ministry of Digital Development and Information.}

\begin{abstract}
Traditional emotional voice conversion (EVC) conditions generation on explicit target emotions like labels or references, defining the target affective state but omitting the direction or nature of the transition. We introduce instruction-guided relative emotional voice conversion, a task where natural-language instructions specify source-conditioned affective transformations (e.g., ``make the speech slightly calmer" or ``sound noticeably more confident") instead of fixed targets. To support this task, we construct TRACE-Instruct, a dataset of relative emotion instructions covering categorical transitions, intensity modifications, and open-ended affective changes. We propose TRACE-EVC, 
a framework built around Emo-Compass, 
a module that models each conversion as a source-anchored rectified flow. 
Rather than conditioning on an explicit target, it predicts the direction and degree of the affective change. Experiments demonstrate that TRACE-EVC accurately follows relative emotion instructions 
while preserving speaker identity, linguistic content, and speech quality, generalizes to unseen speakers,
and remains competitive with conventional EVC systems on standard categorical emotion conversion.
\end{abstract}


\maketitle

\section{INTRODUCTION}

\IEEEPARstart{E}{motional} voice conversion (EVC) aims to modify the emotional expression of speech while preserving speaker identity and linguistic content~\cite{esd}. As an important direction in expressive speech generation, EVC has broad applications in personalized speech interaction, audiobook production, film dubbing, and assistive communication~\cite{EmoInt,mixedevc}. A central challenge in EVC is how to specify the desired emotional transformation. Existing methods address this through label-based control~\cite{seq2seq-EVC,Durflex-EVC}, reference-based control~\cite{sgevc,zest,TextlessEVC,vevo}, and prompt-based control~\cite{Promptevc,ClapFM-EVC}.

\begin{figure}
    \centering
    \includegraphics[width=1.0\linewidth]{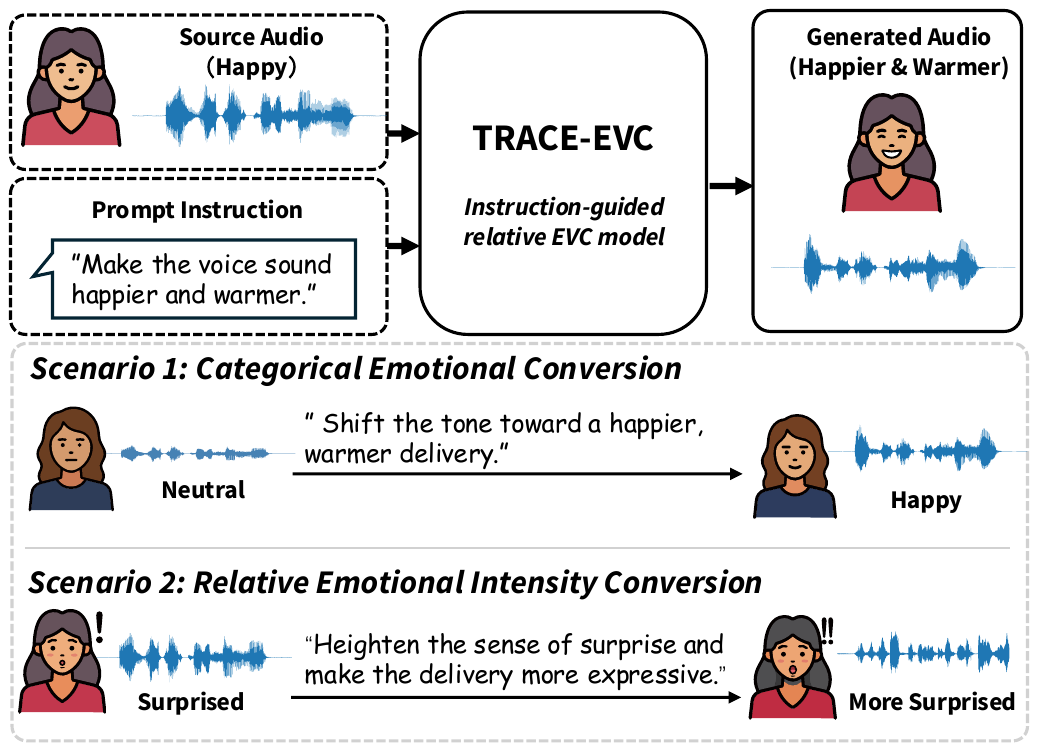}
    \vspace{-5mm}
    \caption{Illustration of the instruction-guided relative EVC task. The instruction specifies how the source emotion should change rather than a fixed target.}
    \vspace{-5mm}
    \label{fig:intro}
\end{figure}

Although these control paradigms have been widely studied, they all require an explicit target condition, such as a target emotion category, a reference utterance that demonstrates the desired emotion, or a description of the final target style. In many practical scenarios, however, such target conditions may be unavailable or difficult to specify in a categorical manner. Instead, an alternative, and arguably more natural specification could be expressed as a relative and directional modification of the source utterance: for example, ``making the speech sound more confident'', ``slightly less excited'', ``warmer'', ``calmer'', or ``more expressive'', while preserving speaker identity and linguistic content. Motivated by this observation, we introduce a new task: instruction-guided relative EVC, in which a natural-language instruction specifies how the emotional expression of a source utterance should be modified, as illustrated in Fig.~\ref{fig:intro}. Unlike conventional EVC settings that convert speech toward a predefined target, this task centers on emotional changes defined relative to the source utterance. Such changes range from categorical emotion shifts to fine-grained modifications of emotional intensity and expressiveness.

This new task introduces two key challenges. First, relative instructions describe affective transformations rather than target emotional states. 
Such transformations may involve changes in valence, arousal, and dominance (VAD), representing emotional pleasantness, activation, and sense of control, respectively, or in prosody rather than a simple emotion-category switch, requiring an interpretable affective representation that captures both the direction and magnitude of emotional change.
Second, existing instruction-style speech datasets describe individual utterances~\cite{textrolspeech}, whereas instruction-guided relative EVC requires descriptions of affective transformations between source and target utterances, leaving a data gap for learning relative emotion instructions.

To address this data gap, we construct TRACE-Instruct, a dataset of relative emotion instructions built from paired emotional speech. Each instruction is generated by prompting a large language model with the source and target style descriptions together with their affective difference, guiding it to describe the affective transformation rather than simply the target style. The resulting instructions provide free-form descriptions of emotional change without explicitly revealing the target emotion labels. Covering cross-emotion shifts, intra-emotion intensity modifications, and open-ended affective transitions, TRACE-Instruct provides the training and evaluation data needed for instruction-guided relative EVC.

To address these challenges, we propose TRACE-EVC, 
a framework for instruction-guided relative EVC. 
Its key component, \textbf{Emo-Compass}, rethinks how prompt-based EVC generates emotion embeddings. Prior work such as PromptEVC~\cite{Promptevc} generates the emotion embedding from a Gaussian noise prior via a diffusion mapper conditioned on the prompt. In our relative setting, however, the source emotion is already given and the instruction is defined relative to it, so we instead anchor the generative process at the source embedding rather than at noise, realizing each conversion as a source-anchored rectified flow whose velocity is exactly the displacement the instruction describes.
A synthesis module then generates the converted speech from the source content, speaker embedding, and predicted emotion embedding, enabling zero-shot conversion without target labels or reference utterances.

Our main contributions are summarized as follows:

\begin{itemize}

\item \textbf{Instruction-guided relative EVC.} We introduce a new EVC task in which natural-language instructions specify source-conditioned affective transformations rather than explicit target emotions, enabling both categorical emotion transitions and fine-grained intensity modification.

\item \textbf{TRACE-Instruct.} We construct TRACE-Instruct, a dataset of relative emotion instructions that supports training and evaluation for instruction-guided relative EVC across categorical transitions, intensity modifications, and open-ended affective transformations.

\item \textbf{TRACE-EVC.} We propose TRACE-EVC, whose Emo-Compass casts relative emotion control as a source-anchored rectified flow, 
enabling EVC without target labels or reference utterances, including for unseen speakers
while supporting flexible instruction following and fine-grained emotion control.

\item \textbf{Experimental validation.} Extensive experiments demonstrate that TRACE-EVC follows relative emotion instructions while preserving speech quality, speaker identity, and linguistic content, and remains competitive with conventional EVC systems on standard categorical emotion conversion.
\end{itemize}

We release the TRACE-Instruct dataset, audio demos, and model code for reproducibility.\footnote{TraceEVC: \url{https://traceevc4-star.github.io/website/}}

\section{Related Work}
\subsection{Emotional Voice Conversion}
Existing EVC methods can be broadly grouped by how the target emotion is specified. Label-based EVC conditions generation on a predefined target emotion category. Prior work improves conversion quality and disentanglement~\cite{seq2seq-EVC,Durflex-EVC}. EVC has also been extended to speaker-independent conversion and zero-shot conversion of unseen speakers~\cite{spk_indenpend,ZSDEVC}.

Reference-based EVC instead specifies the desired emotion through a reference utterance. SGEVC~\cite{sgevc} derives emotion embeddings from reference speech through semi-supervised categorical latent modeling, while ZEST~\cite{zest} performs zero-shot audio-to-audio emotion transfer via speaker disentanglement. TextlessEVC~\cite{TextlessEVC} and HybridVC~\cite{HybridVC} use decomposed, discrete, or multimodal representations for flexible style transfer.

Prompt-based EVC replaces labels and references with natural-language descriptions of the desired style, offering a more flexible interface. PromptEVC~\cite{Promptevc} generates a fine-grained emotion embedding from a noise prior via a diffusion-based prompt mapper conditioned on the style prompt, while ClapFM-EVC~\cite{ClapFM-EVC} and PromptVC~\cite{PromptVC} use natural-language or CLAP-based descriptions to control emotional or stylistic attributes.

Despite these advances, these paradigms remain target-oriented, relying on target emotion labels, reference emotional speech, or target-style descriptions to guide generation. In particular, even prompt-based EVC describes an absolute target style rather than a change from the source. In contrast, our work focuses on source-conditioned emotion transformation, where free-form natural-language instructions describe the desired emotional change relative to the source utterance without requiring target emotion labels or reference speech.

\subsection{Natural-Language-Controlled Speech Generation}

\label{sec:nlc}

Recent work controls speech generation with natural language. Datasets such as TextrolSpeech~\cite{textrolspeech}, PromptTTS~\cite{prompttts}, and SpeechCraft~\cite{speechcraft} pair utterances with descriptions of their speaking style, emotion, or acoustic attributes, and prompt-based models built on such data condition synthesis on these descriptions to control output style. In all cases, however, a description characterizes a single utterance, capturing what it should sound like rather than how it should change. In contrast, our instructions describe the emotional change from a source utterance to a target, providing the supervision for relative control~\cite{instructpix2pix}.

\subsection{Continuous and Fine-Grained Emotion Modeling}
Beyond categorical emotion labels, prior work explores emotion intensity~\cite{seq2seq-EVC,EmoInt}, mixed emotions~\cite{mixedevc}, and dimensional PAD/VAD representations~\cite{PAD}. Most related to ours, EmoSphere-TTS~\cite{emospheretts} represent emotion as a continuous spherical VAD vector for style and intensity control, but still require a target emotion or intensity to be specified.

Although these studies improve fine-grained controllability, they require an explicit control signal to be specified at inference time, such as an intensity label, a mixture weight, or a target VAD coordinate. In contrast, our work grounds a free-form instruction into a continuous VAD space, where VAD serves as an internal representation of the relative change rather than a control value the user must provide.
\begin{figure*}
    \vspace{-5mm}
    \centering
    \includegraphics[width=1.0\linewidth]{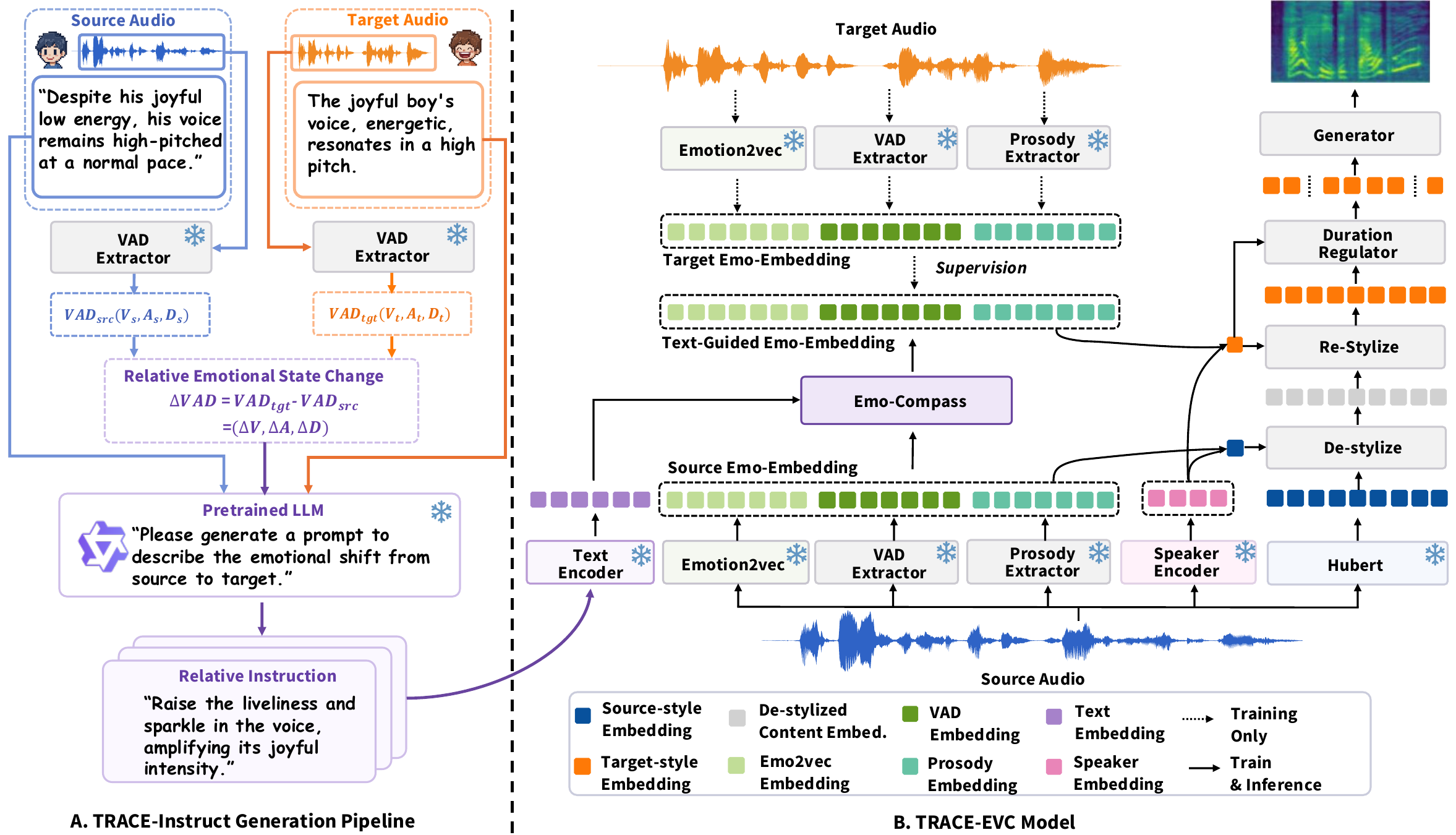}
    \caption{Overview of the proposed framework: a TRACE-Instruct generation pipeline that produces relative instructions, and a TRACE-EVC model that performs the conversion. The target audio supervises Emo-Compass only during training, shown as dashed training-only paths; at inference, the target emotion embedding is predicted from the source and instruction alone, without any target audio.}
    \label{fig:overview}
    \vspace{-5mm}
\end{figure*}

\section{Proposed Framework}
The proposed framework consists of two components, as illustrated in Fig.~\ref{fig:overview}: TRACE-Instruct, which constructs relative emotion instructions from paired emotional speech, and TRACE-EVC, which performs instruction-guided EVC. Given a source utterance and a relative emotion instruction, TRACE-EVC predicts the target emotion embedding corresponding to the instructed transformation and synthesizes the converted speech.

\subsection{TRACE-Instruct Dataset}
\label{sec:dataset}

We construct TRACE-Instruct, an instruction dataset for relative EVC, as shown in Fig.~\ref{fig:overview}(A). 
TRACE-Instruct is built from source-target emotional speech pairs, where each pair provides supervision for generating a natural-language instruction describing the relative emotional change from the source toward the target. Unlike utterance-level descriptions that characterize a single speech sample, TRACE-Instruct focuses on how emotional expression changes from a source utterance toward a desired target state.

\noindent\textbf{Emotional change pair construction.}
 From the parallel ESD corpus~\cite{esd}, we pair recordings of the same speaker and sentence across different emotions to form inter-emotion pairs. From MEAD~\cite{mead}, we form intra-emotion pairs across intensity levels of the same emotion, and open-ended pairs across different emotions at random intensities. Style descriptions for both corpora are taken from TextrolSpeech~\cite{textrolspeech}, denoted $p_{src}$ and $p_{tgt}$.

\noindent\textbf{Relative affective difference.} 
For each pair, a speech emotion recognition (SER) model~\cite{odyssey} predicts the VAD coordinates of the source and target utterances: 

\begin{equation}
\begin{aligned}
\mathbf{VAD}_{src} &= (V_{src}, A_{src}, D_{src}),\\
\mathbf{VAD}_{tgt} &= (V_{tgt}, A_{tgt}, D_{tgt}).
\end{aligned}
\label{eq:vad}
\end{equation}

The relative affective change is then computed as: 
\begin{equation}
\Delta\mathbf{VAD}=\mathbf{VAD}_{tgt}-\mathbf{VAD}_{src}=(\Delta V,\Delta A,\Delta D),
\label{eq:dvad}
\end{equation}

which captures both the direction and magnitude of the emotional transition from the source toward the target. This relative difference provides the LLM with explicit information about how the affective state should change, guiding it to describe the transformation rather than merely the target style.
\noindent\textbf{Instruction generation.} 
Given the source style description $p_{src}$, the target style description $p_{tgt}$, and the relative affective change $\Delta\mathbf{VAD}$, we prompt Qwen3~\cite{qwen3} to generate a set $\mathcal{P}$ of lexically diverse relative emotion instructions:
\begin{equation}
\mathcal{P}=\mathrm{LLM}(p_{src},p_{tgt},\Delta\mathbf{VAD}).
\label{eq:qwen}
\end{equation}
To keep the instructions relative without revealing the target, we apply rule-based filtering that removes explicit target emotion labels, intensity levels, numerical VAD values, and dataset metadata.

\subsection{TRACE-EVC}
\label{sec:trace-evc}
TRACE-EVC consists of two cascaded modules: \textbf{Emo-Compass} and a \textbf{Synthesis Module}, as shown in Fig.~\ref{fig:overview}(B). Given the source emotion embedding and a natural-language instruction, Emo-Compass predicts a target emotion embedding, which the Synthesis Module then uses, together with the source content and speaker representations, to generate the converted speech. Because the speaker is inherited from the source and emotion is conditioned on a continuous embedding rather than a discrete label, TRACE-EVC supports zero-shot conversion of unseen speakers without target labels or reference utterances at inference time.

\noindent\textbf{Emo-Compass.}
Emo-Compass predicts the target emotion embedding $\hat{\mathbf{z}}_{tgt}$ from the source emotion embedding $\mathbf{z}_{src}$ and the instruction embedding $\mathbf{p}$, obtained from a frozen text encoder~\cite{e5}. Each emotion embedding concatenates an emotion2vec~\cite{emo2vec} embedding $\mathbf{e}$, VAD coordinates $\mathbf{a}$, and mean prosodic features $\mathbf{r}$ (mean $F_0$ and energy), giving $\mathbf{z}=[\mathbf{e};\mathbf{a};\mathbf{r}]$. emotion2vec captures category-level emotional information but is less effective at modeling fine-grained intensity variations; we therefore add VAD and prosodic cues to complement it for continuous affective and intensity-related changes. Ablation studies in Sec.~\ref{sec:compass_eval} validate this design.
 
A relative instruction specifies a displacement rather than an absolute endpoint, so the source emotion is a natural starting point. Emo-Compass therefore realizes this source-anchored prediction as a rectified flow in the affective space, whose velocity parameterizes that displacement.
Given the source embedding $\mathbf{z}_{src}$ and the target emotion embedding $\mathbf{z}_{tgt}$, we define a linear trajectory between them as
\begin{equation}
\mathbf{x}(s)=(1-s)\mathbf{z}_{src}+s\mathbf{z}_{tgt},
\label{eq:trajectory}
\end{equation}
where $s$ denotes the flow time. A Transformer $D_\theta$, conditioned on the current state $\mathbf{x}(s)$, the flow time $s$, the source anchor $\mathbf{z}_{src}$, and the instruction embedding $\mathbf{p}$, predicts a velocity $\hat{\mathbf{v}}=D_\theta(\mathbf{x}(s),s;\mathbf{z}_{src},\mathbf{p})$, trained to match the constant target velocity along the trajectory:
\begin{equation}
\mathbf{v}^{*}=\mathbf{z}_{tgt}-\mathbf{z}_{src}.
\label{eq:velocity}
\end{equation}
This target velocity $\mathbf{v}^{*}$ is exactly the relative emotional change from the source toward the target, and is what Emo-Compass learns to predict.
Because each training pair shares the same speaker, the learned displacement $\mathbf{v}^{*}$ captures emotion-induced variation rather than speaker identity, reducing speaker-related leakage.
During inference, the target is unknown, so we integrate the predicted velocity. Starting from $\mathbf{x}^{(0)}=\mathbf{z}_{src}$, we take $K$ Euler steps:
\begin{equation}
\mathbf{x}^{(k+1)}=\mathbf{x}^{(k)}+\tfrac{1}{K}
D_\theta\big(\mathbf{x}^{(k)},\tfrac{k}{K};\mathbf{z}_{src},\mathbf{p}\big),
\qquad \hat{\mathbf{z}}_{tgt}=\mathbf{x}^{(K)}.
\label{eq:infer}
\end{equation}
This yields the text-guided target emotion embedding $\hat{\mathbf{z}}_{tgt}$.

The training objective of Emo-Compass combines a velocity-regression term with four auxiliary guidance terms. The velocity-regression term matches the predicted velocity to the ground-truth displacement $\mathbf{v}^{*}$:
\begin{equation}
\mathcal{L}_{flow}=\mathbb{E}_{s\sim\mathcal{U}[0,1]}\big[\lVert
D_\theta(\mathbf{x}(s),s;\mathbf{z}_{src},\mathbf{p})-\mathbf{v}^{*}\rVert^{2}\big].
\label{eq:flow}
\end{equation}
Since this term alone lets the high-dimensional emotion2vec component dominate the VAD and prosody cues (Sec.~\ref{sec:compass_eval}), we add four guidance terms on the single-step target estimate $\tilde{\mathbf{z}}_{tgt}=\mathbf{x}(s)+(1-s)\hat{\mathbf{v}}=[\tilde{\mathbf{e}};\tilde{\mathbf{a}};\tilde{\mathbf{r}}]$:
\begin{equation}
\left\{
\begin{array}{@{}l@{}}
\mathcal{L}_{cos}=1-\cos(\tilde{\mathbf{e}}-\mathbf{e}_{src},\,\mathbf{e}_{tgt}-\mathbf{e}_{src}),\\
\mathcal{L}_{emo}=\mathrm{CE}(C_\psi(\tilde{\mathbf{e}}),\,y_{tgt}),\\
\mathcal{L}_{aff}=\lVert\tilde{\mathbf{a}}-\mathbf{a}_{tgt}\rVert^2+\lVert\tilde{\mathbf{r}}-\mathbf{r}_{tgt}\rVert^2,\\
\mathcal{L}_{int}=\max(0,\,m-g_i+g_j),
\end{array}
\right.
\label{eq:aux}
\end{equation}
where $C_\psi$ is a frozen emotion classifier, $g_k=g(\tilde{\mathbf{z}}_k)$ an intensity score with margin $m$, and $(i,j)$ MEAD pairs of the same source with intensity $l_i>l_j$. The overall objective is:
\begin{equation}
\mathcal{L}=\mathcal{L}_{flow}+\lambda_{cos}\mathcal{L}_{cos}
+\lambda_{emo}\mathcal{L}_{emo}+\lambda_{int}\mathcal{L}_{int}
+\lambda_{aff}\mathcal{L}_{aff}.
\label{eq:total}
\end{equation}

\noindent\textbf{Synthesis Module.}
The Synthesis Module adapts the DurFlex-EVC~\cite{Durflex-EVC} backbone, replacing its discrete speaker and emotion labels with continuous embeddings to enable zero-shot conversion. It takes the source content from HuBERT~\cite{hubert}, the source speaker embedding $\mathbf{s}$ from EASE~\cite{zest}, and the source and predicted target emotion embeddings $\mathbf{z}_{src}$ and $\hat{\mathbf{z}}_{tgt}$.

The speaker and emotion embeddings are first combined into source and target style representations through an emotion projection $\phi_{emo}$:
\begin{equation}
\mathbf{m}_{src}=\mathbf{s}+\phi_{emo}(\mathbf{z}_{src}),\qquad
\mathbf{m}_{tgt}=\mathbf{s}+\phi_{emo}(\hat{\mathbf{z}}_{tgt}).
\label{eq:style}
\end{equation}
The two style vectors share the speaker term $\mathbf{s}$ and differ only in emotion: $\mathbf{m}_{src}$ de-stylizes the content and $\mathbf{m}_{tgt}$ re-stylizes it, converting emotion while preserving content and speaker identity.

Following DurFlex-EVC~\cite{Durflex-EVC}, the re-stylized content is decoded by a conditional diffusion decoder that generates the mel-spectrogram by denoising from Gaussian noise. With $\mathbf{y}_t$ the mel-spectrogram noised with Gaussian noise $\boldsymbol{\epsilon}$ at diffusion step $t$, and $\mathbf{c}$ the frame-level content condition, the decoder $\boldsymbol{\epsilon}_\theta$ is trained to predict $\boldsymbol{\epsilon}$:
\begin{equation}
\mathcal{L}_{mel}=\mathbb{E}_{t,\boldsymbol{\epsilon}}
\big[\lVert\boldsymbol{\epsilon}_\theta(\mathbf{y}_t,t\mid\mathbf{c},
\mathbf{m}_{tgt})-\boldsymbol{\epsilon}\rVert_{2}^{2}\big].
\label{eq:mel}
\end{equation}
Conditioned on the emotion embedding predicted by Emo-Compass, the Synthesis Module is trained by self-reconstruction with $\mathcal{L}_{syn}=\mathcal{L}_{mel}+\lambda_{unit}\mathcal{L}_{unit}+\lambda_{mp}\mathcal{L}_{mp}$, where $\mathcal{L}_{unit}$ is a cross-entropy on HuBERT units and $\mathcal{L}_{mp}$ an MSE on utterance-level mean $F_0$ and energy. A pretrained HiFi-GAN~\cite{hifigan} vocoder converts the generated mel-spectrogram into waveform.

\section{Experiments and Results}

\subsection{Experimental Setup}
\noindent\textbf{Datasets.}
We conduct experiments on ESD~\cite{esd} and MEAD~\cite{mead}. For inter-emotion conversion, we use the English portion of ESD, which contains five emotions (Neutral, Happy, Angry, Sad, and Surprise), 17,500 utterances, and approximately 11.9 hours of speech. Following the standard protocol, we hold out 30 utterances per emotion for each speaker for testing, yielding a seen-speaker setting for comparison with prior EVC baselines. For intra-emotion intensity conversion, we use MEAD, which contains four emotions with three intensity levels. We use 36 training, 3 validation, and 4 unseen test actors for zero-shot evaluation, and additionally evaluate a seen-speaker setting using held-out utterances from the training actors. After retaining paired recordings across intensity levels, the subset contains 12,350 utterances and approximately 13.8 hours of speech. 
In both datasets, ground-truth source--target pairs are formed within each split, and their instructions follow the same split.
All ESD experiments thus use seen speakers for a fair comparison with label-based baselines, while zero-shot speaker generalization is evaluated only on unseen MEAD speakers.

\noindent\textbf{Baselines.}
We compare TRACE-EVC with three groups of baselines that differ in how they handle relative instructions: dedicated EVC systems use absolute target conditions, LLM-mediated EVC systems convert relative instructions into target labels, and direct LALMs generate speech directly from the source audio and instruction. TRACE-EVC is specifically designed for relative EVC.

\begin{itemize}
    \item \textbf{Dedicated EVC baselines.}
    Reference-based baselines include the original SGEVC~\cite{sgevc}, ZEST~\cite{zest}, and VEVO~\cite{vevo}, while label-based baselines include our label-controlled SGEVC and DurFlex-EVC~\cite{Durflex-EVC}. For SGEVC, we implement label-based control by directly specifying the target emotion category.
    \item \textbf{LLM-based instruction-to-label EVC baselines.}
    To support relative instructions, an SER model~\cite{emo2vec} first estimates the source emotion, and an LLM~\cite{qwen3} combines it with the instruction to predict a target emotion label for label-controlled SGEVC or DurFlex-EVC. These systems therefore handle relative control indirectly through target-label prediction.
    \item \textbf{Direct LALM baselines.}
    We evaluate MiniCPM-o 4.5~\cite{minicpm}, Qwen3-Omni-30B~\cite{qwen3-omni}, and Kimi-Audio~\cite{kimi} for zero-shot relative EVC. Each model directly takes the source audio and relative instruction as input. We additionally provide a job description (JD) specifying the task (Fig.~\ref{fig:jd_prompt}) to reduce task-related hallucinations. Table~\ref{tab:jd_behavior} compares their behavior and WER with and without the JD.
\end{itemize}
We exclude prompt-based EVC models such as PromptEVC~\cite{Promptevc} because no public implementation is available and their prompts specify absolute target styles rather than relative affective changes.

\noindent\textbf{Objective Metrics.}
All objective metrics are averaged over the test set. We evaluate intelligibility using WER from Whisper-large-v3~\cite{whisper}. To avoid circular evaluation, emotion similarity is measured with MERaLiON-SER-v1~\cite{MERaLiON-SER}, independent of the Odyssey SER used for dataset construction and emotion2vec~\cite{emo2vec} used for intermediate emotion representations. EECS is the cosine similarity between emotion embeddings of the converted speech and ground-truth target, with higher values indicating better emotion alignment. We additionally report emotion classification accuracy ($\mathrm{ACC}_{\mathrm{cls}}$) using a wav2vec2-XLSR~\cite{wav2vec} classifier fine-tuned on ESD. Speaker similarity (SECS) is computed between Resemblyzer~\cite{ge2e} embeddings of the converted speech and target recording. Speech quality is measured using UTMOS~\cite{utmos} on ESD and NISQA~\cite{nisqa} on MEAD.
\noindent\textbf{Subjective Evaluation.}
We recruit 15 participants with normal hearing and English proficiency to rate naturalness (nMOS) and instruction following (MOS-I) on a 1--5 scale. For nMOS, all systems are evaluated on the same 8 conversion pairs from the test set. For MOS-I, we adapt the instruction-following MOS protocol~\cite{voxinstruct} to relative EVC and evaluate 10 randomly selected converted utterances per setting. Raters are presented with the source utterance, converted utterances, and instruction, and score how well the output follows the instructed relative change. This paired presentation focuses the evaluation on relative rather than absolute emotion. Each sample is rated by all 15 participants, and we report the mean.

\noindent\textbf{Implementation Details.}
The natural language instruction is encoded by a frozen E5~\cite{e5} encoder, and content is represented by HuBERT~\cite{hubert} discrete units over a $200$-cluster codebook.
Emo-Compass is a $6$-layer Transformer trained as a source-anchored rectified flow with $(\lambda_{cos},\lambda_{emo},\lambda_{int},\lambda_{aff})=(0.1,0.2,0.5,1.0)$, with $5$ Euler steps at inference.
The two modules are trained separately with AdamW at a learning rate of $1\times10^{-4}$: Emo-Compass for $100$ epochs and the synthesis module for about $40$k steps with $(\lambda_{unit},\lambda_{mp})=(0.1,2.0)$, on $4$ A100 GPUs. Mel-spectrograms ($80$ bins, $16$\,kHz, hop size $256$) are converted to waveforms by a HiFi-GAN~\cite{hifigan} pretrained on LibriTTS~\cite{libritts}, without fine-tuning.
\begin{table*}[t]
\centering
\small
\setlength{\tabcolsep}{3pt}
\renewcommand{\arraystretch}{0.8}
\caption{Comparison with EVC and LALM baselines. SGEVC-L denotes our label-controlled variant of SGEVC; NLP denotes a free-form natural-language prompt, and JD an additional task description. Params reports the core generative model size, excluding frozen modules. EECS, SECS, and UTMOS are decimals; $\mathrm{ACC}_{\mathrm{cls}}$ and WER are percentages; nMOS is reported with 95\% CI. SECS is omitted for LALMs using default speakers. VEVO and LALMs serve as large-scale references. Best and second-best results are bolded and underlined.}
\label{tab:evc_baselines}
\begin{tabular}{llrccccccc}
\toprule
\textbf{Model} & \textbf{Control Input} & \textbf{Params} & \textbf{Training Data (h)}
& \textbf{WER}$\downarrow$
& \textbf{EECS}$\uparrow$
& $\mathbf{ACC}_{\mathrm{cls}}\uparrow$
& \textbf{SECS}$\uparrow$
& \textbf{UTMOS}$\uparrow$
& \textbf{nMOS}$\uparrow$ \\
\midrule
\rowcolor{gray!15}
\multicolumn{10}{l}{\textbf{Dedicated EVC baselines}} \\
SGEVC-L & Emotion Label & 40.8M & 11.9 & 9.81 & 0.73 & 80.44 & 0.60 & 3.23 & 3.84 $\pm$ 0.18 \\
DurFlex-EVC & Emotion Label & 158.7M & 11.9 & 10.05 & \underline{0.81} & \underline{89.50} & 0.56 & 3.17 & 3.67 $\pm$ 0.23 \\
SGEVC & Speech Reference & 40.8M & 11.9 & 9.35 & 0.66 & 72.60 & 0.59 & 3.26 & 3.72 $\pm$ 0.20 \\
ZEST & Speech Reference & 333.0M & 11.9 & \underline{8.09} & 0.79 & 78.58 & 0.55 & 3.16 & 3.36 $\pm$ 0.21 \\
VEVO & Speech Reference & 1.12B & 0.1M & 8.48 & 0.52 & 32.42 & 0.64 & 3.61 & 3.87 $\pm$ 0.13\\
\midrule
\rowcolor{gray!15}
\multicolumn{10}{l}{\textbf{LLM-based instruction-to-label EVC baselines}} \\
SGEVC-L & LLM Predicted Label & 40.8M & 11.9 & 9.77 & 0.70 & 74.12 & 0.58 & 3.23 & 3.80 $\pm$ 0.21 \\
DurFlex-EVC & LLM Predicted Label & 158.7M & 11.9 & 10.04 & 0.78 & 82.79 & 0.57 & 3.21 & 3.69 $\pm$ 0.17 \\
\midrule
\rowcolor{gray!15}
\multicolumn{10}{l}{\textbf{Direct LALM baselines}} \\
MiniCPM-o 4.5 & JD + NLP & 9.37B & -- & \textbf{6.41} & 0.02 & 10.77 & \textbf{0.70} & \underline{4.13} & 3.88 $\pm$ 0.28 \\
Qwen3-Omni-30B & JD + NLP & 35.3B & 20M & 8.52 & 0.17 & 25.21 & -- & \textbf{4.32} & 3.85 $\pm$ 0.21 \\
Kimi-Audio & JD + NLP & 9.77B & 13M & 19.82 & 0.15 & 25.42 & -- & 3.26 & \textbf{4.06 $\pm$ 0.23} \\
\midrule
\rowcolor{gray!15}
\multicolumn{10}{l}{\textbf{Proposed method}} \\
\textbf{TRACE-EVC (Ours)} & NLP & 205.7M & 11.9 & 9.96 & \textbf{0.85} & \textbf{93.00} & \underline{0.68} & 3.57 & \underline{3.98 $\pm$ 0.11} \\
\quad w/o Emo-Compass & NLP & 159.0M & 11.9 & 10.90 & 0.65 & 43.08 & 0.66 & 3.38 & -- \\
\bottomrule
\vspace{-6mm}
\end{tabular}
\end{table*}
\begin{table}[t]
\centering
\caption{LALM output behavior and WER with and without JD. All values are percentages. EVC denotes exact or approximate reproduction of source content, regardless of emotional accuracy. Replies include responses and other outputs.}
\label{tab:jd_behavior}
\small
\setlength{\tabcolsep}{2pt}
\renewcommand{\arraystretch}{0.8}
\begin{tabular*}{\columnwidth}{@{}l@{\extracolsep{\fill}}llllll@{}}
\toprule
\multirow{2}{*}{Model} &
\multirow{2}{*}{JD} &
\multirow{2}{*}{WER} &
\multirow{2}{*}{EVC} &
Adds &
Reads &
\multirow{2}{*}{Replies} \\
& & & & content & prompt & \\
\midrule
\multirow{2}{*}{MiniCPM-o 4.5}
& No  & 13.37\% & 91\% & 2\% & 3\% & 3\% \\
& Yes & 6.41\%  & 94\% & 3\% & 0\% & 4\% \\
\midrule
\multirow{2}{*}{\shortstack[l]{Qwen3-Omni-\\30B}}
& No  & 226.26\% & 73\% & 9\% & 6\% & 13\% \\
& Yes & 8.52\%   & 93\% & 2\% & 0\% & 5\% \\
\midrule
\multirow{2}{*}{Kimi-Audio}
& No  & 239.72\% & 11\% & 2\% & 0.4\% & 86\% \\
& Yes & 19.82\%  & 89\% & 2\% & 1\% & 9\% \\
\bottomrule
\vspace{-6mm}
\end{tabular*}
\end{table}
\begin{figure}[t]
\centering
\begin{tcolorbox}[
    width=\columnwidth,
    colback=gray!10,
    colframe=black,
    colbacktitle=black,
    coltitle=white,
    title={Job Description (JD)},
    fonttitle=\bfseries\sffamily\small,
    fontupper=\small,
    sharp corners=south,
    boxrule=0.5pt,
    arc=2mm,
    top=2mm,
    bottom=2mm,
    left=2mm,
    right=2mm,
    before skip=0pt,
    after skip=0pt
]
Listen to the speech above. Say exactly the same sentence again,
word for word, keeping the same speaker's voice, but change the
way it is delivered as follows: \texttt{\{prompt\}}
Do not add, remove or change any words, and do not reply to or
comment on the content.

\end{tcolorbox}
\caption{Job description (JD) template.
\texttt{\{prompt\}} denotes the relative emotion instruction.}
\label{fig:jd_prompt}
\vspace{-3mm}
\end{figure}
\subsection{Results on EVC}


\noindent\textbf{Inter-emotion.} Dedicated EVC baselines are given explicit target labels or reference speech, while the instruction-to-label baselines must infer the target emotion from the source and relative instruction. As shown in Table~\ref{tab:evc_baselines}, replacing explicit label control with the SER--LLM pipeline reduces both EECS and $\mathrm{ACC}_{\mathrm{cls}}$ for SGEVC-L and DurFlex-EVC, illustrating the additional challenge of relative instruction following.

Direct LALMs achieve competitive WER and UTMOS but substantially lower emotion alignment. Adding the JD improves source-content reproduction and reduces WER for all three LALMs (Table~\ref{tab:jd_behavior}), although non-EVC responses and added content remain. TRACE-EVC achieves the highest EECS and $\mathrm{ACC}_{\mathrm{cls}}$, while maintaining competitive intelligibility and speech quality. It also achieves higher SECS than all dedicated and instruction-to-label EVC baselines, second only to MiniCPM-o 4.5. Removing Emo-Compass substantially degrades emotion conversion, supporting explicit prediction of the instruction-guided target emotion embedding.

\noindent\textbf{Intra-emotion.}
We evaluate relative intensity conversion on MEAD under seen- and unseen-speaker conditions. As shown in Table~\ref{tab:e2e_mead}, several LALMs obtain higher NISQA or lower WER, while TRACE-EVC achieves the highest EECS on unseen speakers and the highest reported SECS in both settings. On seen speakers, MiniCPM-o 4.5 obtains slightly higher EECS. These results show that TRACE-EVC maintains strong target-emotion alignment and speaker preservation when generalizing to unseen speakers. Removing Emo-Compass consistently degrades all four metrics.

\begin{figure}
    \centering
    \includegraphics[width=0.95\linewidth]{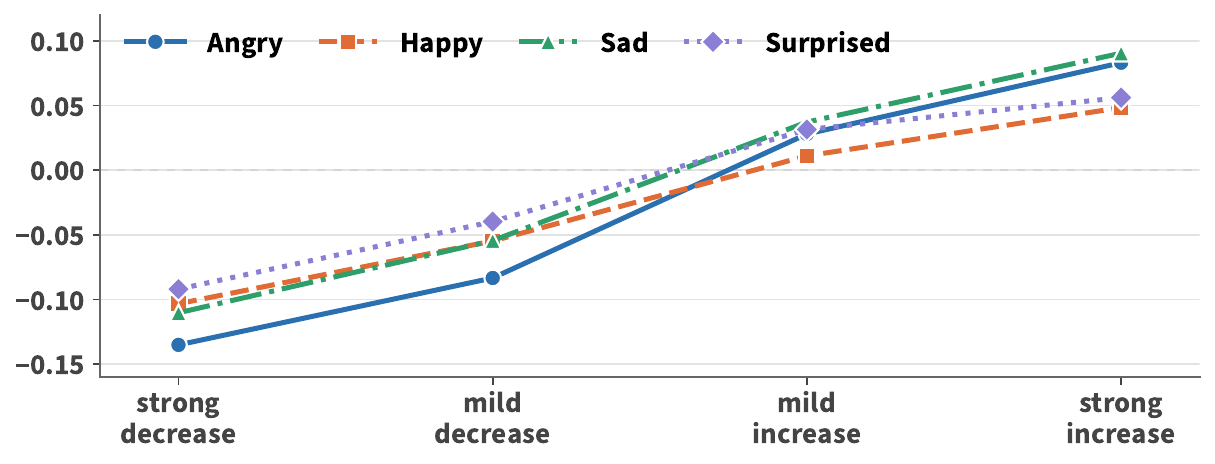}
    \caption{Mean arousal shift under relative intensity instructions on MEAD. Free-form instructions are grouped by direction and strength, from strong decrease to strong increase. Arousal shifts increase monotonically across all emotions.}
    \label{fig:arousal}
    \vspace{-2mm}
\end{figure}

Fig.~\ref{fig:arousal} further shows that mean arousal shifts increase monotonically from strong decreases to strong increases across all four emotions, demonstrating graded relative intensity control.

\begin{table}[t]
\centering
\caption{Intra-emotion conversion on MEAD. SECS and EECS are decimals, and WER is a percentage. SECS is omitted for Qwen3-Omni and Kimi-Audio because they use default speakers. Bold values indicate the best conversion result in each speaker setting, excluding GT.}
\renewcommand{\arraystretch}{0.8}
\resizebox{1.0\columnwidth}{!}{
\label{tab:e2e_mead}
\setlength{\tabcolsep}{4pt}
\begin{tabular}{lcccc}
\toprule
System & NISQA $\uparrow$ & SECS $\uparrow$ & EECS $\uparrow$ & WER $\downarrow$ \\
\midrule
\rowcolor{gray!15}
\multicolumn{5}{l}{\textbf{Unseen speakers}} \\
GT                   & 3.248 & 1.0 & 1.0 & 8.6 \\
MiniCPM-o 4.5        & 3.984 & 0.678 & 0.905 & 14.0 \\
Qwen3-Omni-30B       & \textbf{4.801} & -- & 0.821 & \textbf{8.8} \\
Kimi-Audio           & 4.083 & -- & 0.742 & 18.9 \\
\textbf{TRACE-EVC (Ours)}        & 3.147 & \textbf{0.706} & \textbf{0.925} & 14.8 \\
\quad w/o Emo-Compass & 3.011 & 0.698 & 0.894 & 16.6 \\
\midrule
\rowcolor{gray!15}
\multicolumn{5}{l}{\textbf{Seen speakers}} \\
GT                   & 3.366 & 1.0 & 1.0 & 11.9 \\
MiniCPM-o 4.5        & 3.983 & 0.692 & \textbf{0.895} & 9.9 \\
Qwen3-Omni-30B       & \textbf{4.778} & -- & 0.810 & \textbf{6.4} \\
Kimi-Audio           & 4.088 & -- & 0.738 & 19.4 \\
\textbf{TRACE-EVC (Ours)}        & 3.193 & \textbf{0.831} & 0.862 & 16.5 \\
\quad w/o Emo-Compass & 3.052 & 0.802 & 0.844 & 18.4 \\
\bottomrule

\end{tabular}}
\end{table}

\noindent\textbf{Instruction following.}
As shown in Table~\ref{tab:if}, TRACE-EVC achieves higher MOS-I scores than all three LALM baselines, MiniCPM-o 4.5, Qwen3-Omni-30B, and Kimi-Audio, in both inter- and intra-emotion settings. For inter-emotion conversion, the higher MOS-I indicates more accurate realization of the instructed transition from the source emotion toward a different emotion. For intra-emotion conversion, it reflects better adherence to instructed intensity changes relative to the source. Together, these results provide perceptual evidence that TRACE-EVC follows relative instructions across both categorical and fine-grained emotion changes.

\begin{table}[t]
\centering
\caption{Instruction-following MOS (MOS-I) scores for inter- and intra-emotion conversion, rated on a 1--5 scale (mean $\pm$ 95\% confidence interval). Higher is better.}
\renewcommand{\arraystretch}{1}
\label{tab:if}
\small
\setlength{\tabcolsep}{5pt}
\begin{tabular}{lcc}
\toprule
System
& \shortstack{Inter-emotion\\(ESD)}
& \shortstack{Intra-emotion\\(MEAD)} \\
\midrule
Ground truth      & 4.21 $\pm$ 0.12 & 4.12 $\pm$ 0.13\\
\hline
MiniCPM-o 4.5     & 2.85 $\pm$ 0.21 & 2.86 $\pm$ 0.21 \\
Qwen3-Omni-30B    & 3.06 $\pm$ 0.21 & 3.36 $\pm$ 0.19 \\
Kimi-Audio        & 2.64 $\pm$ 0.20 & 3.02 $\pm$ 0.17 \\
\textbf{TRACE-EVC (Ours)}  & \textbf{3.83 $\pm$ 0.13} & \textbf{3.96 $\pm$ 0.14} \\
\bottomrule
\vspace{-4mm}
\end{tabular}
\end{table}

\noindent\textbf{Ordered emotional control.}
We further evaluate ordered control of valence and arousal using 10-level relative instructions generated by Claude Opus 5\footnote{One ten-level instruction sequence and its corresponding generated speech samples are provided on our demo page.}
, specifying varying degrees of increase or decrease in each attribute. We apply these instructions to 25 ESD source utterances (five per emotion) and extract valence and arousal scores from the generated speech using MERaLiON-SER-v1. Following the cross-modality emotion ordinality test of EmotionRankCLAP~\cite{emorank}, we measure how well the generated attributes preserve the instruction ordering: Kendall's $\tau$~\cite{tau} is computed per source and averaged over all 25 sources.

Table~\ref{tab:vad_control_levels} reports results under both the full 10-level ordering and a binary increase/decrease ordering. TRACE-EVC achieves the highest Kendall's $\tau$ for both valence and arousal in both settings, indicating stronger correspondence between instruction strength and the resulting affective change than the evaluated LALMs.

\begin{table}[t]
\centering
\small
\setlength{\tabcolsep}{5pt}
\renewcommand{\arraystretch}{0.9}
\caption{Ordered instruction-following performance for valence and arousal. Kendall's $\tau$ is computed per source and averaged over 25 sources under the full ten-level ordering and a binary increase/decrease ordering. Bold values indicate the best result in each column.}
\label{tab:vad_control_levels}
\begin{tabular*}{\columnwidth}{@{\extracolsep{\fill}}lcccc@{}}
\toprule
& \multicolumn{2}{c}{Ten-level $\tau\uparrow$}
& \multicolumn{2}{c}{Up/down $\tau\uparrow$} \\
\cmidrule(lr){2-3}\cmidrule(lr){4-5}
\textbf{Model}
& Valence & Arousal
& Valence & Arousal \\
\midrule
Kimi-Audio    & 0.132  & 0.216 & 0.140  & 0.247 \\
Qwen3-Omni    & -0.004 & 0.036 & -0.027 & 0.001 \\
MiniCPM-o 4.5 & 0.074  & 0.282 & 0.097  & 0.326 \\
\midrule
\textbf{TRACE-EVC (Ours)} & \textbf{0.253} & \textbf{0.550} & \textbf{0.369} & \textbf{0.667} \\
\bottomrule
\end{tabular*}
\end{table}

\subsection{Emo-Compass Evaluation}
\label{sec:compass_eval}
\begin{table}[t]
\centering

\caption{Intrinsic evaluation and cue ablation of Emo-Compass. \textit{emo\_acc}, EECS, and VAD L2 are reported as decimals, and \textit{int\_MSE} as mean squared error.}
\renewcommand{\arraystretch}{0.8}
\label{tab:compass_eval}
\setlength{\tabcolsep}{4pt}
\begin{tabular}{lcccr}
\toprule
System & emo\_acc $\uparrow$ & EECS $\uparrow$ & VAD L2 $\downarrow$ & int\_MSE $\downarrow$ \\
\midrule
\rowcolor{gray!15}
\multicolumn{5}{l}{\textbf{ESD}} \\
Emo-Compass & 0.994 & 0.949 & \textbf{0.125} & -- \\
\quad w/o VAD             & \textbf{0.996} & \textbf{0.953} & 0.159 & -- \\
\quad w/o Prosody        & 0.995 & 0.944 & 0.130 & -- \\
\quad w/o VAD \& Prosody & 0.995 & 0.952 & 0.160 & --\\
Random      & 0.204 & 0.491 & 0.315 & -- \\
Regression &0.918 & 0.879& 0.135 & --\\
\midrule
\rowcolor{gray!15}
\multicolumn{5}{l}{\textbf{MEAD}} \\
Emo-Compass              & 0.991 & 0.965 & \textbf{0.115} & \textbf{7.15} \\
\quad w/o VAD            & \textbf{0.997}& \textbf{0.969} & 0.191 & 14.13 \\
\quad w/o Prosody        & 0.993 & 0.964 & 0.121 & 9.85 \\
\quad w/o VAD \& Prosody & 0.992 & 0.966 & 0.190 & 14.98 \\
Random                   & 0.244 & 0.586 & 0.281 & 28.79 \\
Regression               &0.901 & 0.893& 0.124 & 9.72\\
\bottomrule
\vspace{-8mm}
\end{tabular}
\end{table}

We evaluate Emo-Compass using four component-level metrics: emotion classification accuracy (\textit{emo\_acc}), EECS between the predicted and ground-truth target emotion embeddings, VAD normalized euclidean distance (\textit{VAD L2}) between their VAD coordinates, and intensity MSE (\textit{int\_MSE}). Unlike the end-to-end $\mathrm{ACC}_{\mathrm{cls}}$ in Table~\ref{tab:evc_baselines}, \textit{emo\_acc} applies an emotion2vec-based classifier directly to the predicted target emotion embedding $\hat{z}_{tgt}$ rather than to synthesized audio.

We compare Emo-Compass with two baselines. \textit{Random} shuffles target emotion embeddings across test pairs, preserving the target marginal distribution while breaking the source--instruction--target correspondence. \textit{Regression} directly predicts the target emotion embedding from the same source embedding and instruction.

As shown in Table~\ref{tab:compass_eval}, Emo-Compass substantially outperforms the random permutation baseline on both ESD and the unseen-speaker MEAD split. This indicates that Emo-Compass learns instruction-conditioned affective changes rather than relying on dataset-level target statistics.
The ablation results highlight the complementary roles of the affective components. Across all variants, the emotion2vec representation supports high emotion classification accuracy and EECS, while the full model achieves the lowest VAD and intensity prediction errors. Removing the VAD cues increases both VAD L2 and \textit{int\_MSE}, showing that explicit VAD cues are important for learning continuous intensity changes. Prosody ablation also increases \textit{int\_MSE} on MEAD, indicating that prosodic information contributes to intensity prediction. Together, these results support combining emotion representations, explicit VAD cues, and prosodic information for fine-grained affective modeling.


The comparison with direct regression further evaluates the choice of transformation modeling. Both methods receive the same source embedding and instruction, but the regression baseline directly predicts the target embedding, whereas Emo-Compass learns a source-anchored velocity field and integrates it to obtain the target. Emo-Compass achieves higher emotion classification accuracy and EECS, lower VAD error on both datasets and lower intensity MSE on MEAD. These results support the source-anchored flow formulation over the evaluated direct regression baseline for 
modeling both categorical and continuous affective changes.
The improvements on the unseen-speaker MEAD split further suggest that this formulation remains effective when the source speaker is absent from training.

\subsection{TRACE-Instruct Analysis}
\noindent\textbf{Instruction diversity.} As shown in Table~\ref{tab:prompt_diversity}, TRACE-Instruct attains a slightly higher Hypergeometric Distribution D (HD-D)~\cite{hdd} than TextrolSpeech~\cite{textrolspeech}, indicating marginally greater lexical diversity. Its low Self-BLEU~\cite{zhu2018texygen} further suggests limited redundancy among the generated instructions, whereas per-pair Self-BLEU is undefined for single-prompt TextrolSpeech. In terms of instruction length, TRACE-Instruct produces consistently longer and more descriptive instructions. Overall, it offers lexically diverse, low-redundancy descriptions of emotional transitions for instruction-guided emotion control.

\noindent\textbf{Instruction alignment.}
The high MOS-I scores of the ground-truth target recordings in Table~\ref{tab:if} further suggest that the instructions accurately describe the emotional changes between the paired recordings, supporting the quality of the relative emotion annotations.

\begin{table}[h!]
\caption{Instruction diversity. TRACE-Instruct results are averaged over ESD and MEAD; per-pair S-BLEU is N/A for single-prompt TextrolSpeech. Instruction length is measured in words.} 
\label{tab:prompt_diversity}
\centering
\small
\setlength{\tabcolsep}{3pt}
\begin{tabular}{@{}l c c c c c@{}}
\toprule
\multirow{2}{*}{\textbf{Dataset}}
& \multirow{2}{*}{\textbf{\#Prompts}}
& \multirow{2}{*}{\textbf{HD-D}$\uparrow$}
& \multirow{2}{*}{\textbf{S-BLEU}$\downarrow$}
& \multicolumn{2}{c}{\textbf{Instruction Len.}} \\
\cmidrule(lr){5-6}
& & & & \textbf{Mean} & \textbf{Std} \\
\midrule
TRACE-Instruct & \textbf{0.54M} & \textbf{0.776} & \textbf{0.028} & 16.99 & 4.67 \\
TextrolSpeech  & 0.24M & 0.774 & -- & 12.64 & 3.97 \\
\bottomrule
\vspace{-6mm}
\end{tabular}
\end{table}
\section{Limitations}
This work focuses on utterance-level emotion control, where instructions specify an overall affective change relative to the source speech. Finer temporal control, such as varying emotional expression within an utterance, remains a direction for future work. Our evaluation is limited to English speech from ESD and MEAD, whose paired recordings enable controlled comparison across emotions and intensity levels. Generalization to spontaneous speech, other languages, and less controlled recording conditions has not yet been evaluated. Extending TRACE-Instruct to these settings will require suitable paired speech data and compatible affective annotation models.

Preserving speaker identity while modifying emotional expression also raises potential risks of unauthorized impersonation and misrepresentation of emotional intent. We therefore recommend use with speaker consent and clear disclosure of synthetic or modified speech.

\section{Conclusion}
We introduced instruction-guided relative EVC, where natural-language instructions describe emotional changes relative to the source rather than a fixed target, and built TRACE-Instruct to support it across inter-emotion, intensity, and open-ended changes. 
TRACE-EVC models
each instruction as a source-anchored trajectory in a continuous affective space, following relative instructions while preserving quality, speaker similarity, and intelligibility. It outperforms much larger audio-language models on relative instruction following, both in perceptual ratings and in ordered control of valence and arousal, while remaining competitive with conventional EVC baselines that are given explicit target conditions.

\newpage
\bibliographystyle{IEEEtran}
\bibliography{TRACE}

@IEEEtranBSTCTL{IEEEexample:BSTcontrol,
  CTLuse_forced_etal       = "yes",
  CTLmax_names_forced_etal = "6",
  CTLnames_show_etal       = "1"
}

@article{esd,
  title={Emotional voice conversion: Theory, databases and esd},
  author={Zhou, Kun and Sisman, Berrak and Liu, Rui and Li, Haizhou},
  journal={Speech Communication},
  volume={137},
  pages={1--18},
  year={2022},
  publisher={Elsevier}
}

@ARTICLE{EmoInt,
  author={Zhou, Kun and Sisman, Berrak and Rana, Rajib and Schuller, Bj{\"o}rn W. and Li, Haizhou},
  journal={IEEE Transactions on Affective Computing}, 
  title={Emotion Intensity and its Control for Emotional Voice Conversion}, 
  year={2023},
  volume={14},
  number={1},
  pages={31--48},
  doi={10.1109/TAFFC.2022.3175578}}

@inproceedings{mixedevc,
  title     = {{Mixed-EVC: Mixed Emotion Synthesis and Control in Voice Conversion}},
  author    = {Kun Zhou and Berrak Sisman and Carlos Busso and Bin Ma and Haizhou Li},
  year      = {2024},
  booktitle = {{The Speaker and Language Recognition Workshop (Odyssey 2024)}},
  pages     = {180--186},
  doi       = {10.21437/odyssey.2024-26},
}

@inproceedings{seq2seq-EVC,
  title     = {{Limited Data Emotional Voice Conversion Leveraging Text-to-Speech: Two-Stage Sequence-to-Sequence Training}},
  author    = {Kun Zhou and Berrak Sisman and Haizhou Li},
  year      = {2021},
  booktitle = {{Interspeech 2021}},
  pages     = {811--815},
  doi       = {10.21437/Interspeech.2021-781},
  issn      = {2958-1796},
}

@ARTICLE{Durflex-EVC,
  author={Oh, Hyung-Seok and Lee, Sang-Hoon and Cho, Deok-Hyeon and Lee, Seong-Whan},
  journal={IEEE Transactions on Affective Computing}, 
  title={DurFlex-EVC: Duration-Flexible Emotional Voice Conversion Leveraging Discrete Representations Without Text Alignment}, 
  year={2025},
  volume={16},
  number={3},
  pages={1660--1674},
  doi={10.1109/TAFFC.2025.3530920}}

@inproceedings{sgevc,
  title     = {{Emotional Voice Conversion with Semi-Supervised Generative Modeling}},
  author    = {Hai Zhu and Huayi Zhan and Hong Cheng and Ying Wu},
  year      = {2023},
  booktitle = {{Interspeech 2023}},
  pages     = {2278--2282},
  doi       = {10.21437/Interspeech.2023-251},
  issn      = {2958-1796},
}

@INPROCEEDINGS{zest,
  author={Dutta, Soumya and Ganapathy, Sriram},
  booktitle={ICASSP 2024 - 2024 IEEE International Conference on Acoustics, Speech and Signal Processing (ICASSP)}, 
  title={Zero Shot Audio To Audio Emotion Transfer With Speaker Disentanglement}, 
  year={2024},
  volume={},
  number={},
  pages={10371--10375},
  doi={10.1109/ICASSP48485.2024.10445962}}

@inproceedings{vevo,
  author       = {Xueyao Zhang and Xiaohui Zhang and Kainan Peng and Zhenyu Tang and Vimal Manohar and Yingru Liu and Jeff Hwang and Dangna Li and Yuhao Wang and Julian Chan and Yuan Huang and Zhizheng Wu and Mingbo Ma},
  title        = {Vevo: Controllable Zero-Shot Voice Imitation with Self-Supervised Disentanglement},
  booktitle    = {{ICLR}},
  publisher    = {OpenReview.net},
  year         = {2025}
}

@inproceedings{Promptevc,
  title     = {{PromptEVC: Controllable Emotional Voice Conversion with Natural Language Prompts}},
  author    = {Tianhua Qi and Shiyan Wang and Cheng Lu and Tengfei Song and Hao Yang and Zhanglin Wu and Wenming Zheng},
  year      = {2025},
  booktitle = {{Interspeech 2025}},
  pages     = {4588--4592},
  doi       = {10.21437/Interspeech.2025-948},
  issn      = {2958-1796},
}

@inproceedings{TextlessEVC,
    title = "Textless Speech Emotion Conversion using Discrete {\&} Decomposed Representations",
    author = "Kreuk, Felix  and
      Polyak, Adam  and
      Copet, Jade  and
      Kharitonov, Eugene  and
      Nguyen, Tu Anh  and
      Rivi{\`e}re, Morgan  and
      Hsu, Wei-Ning  and
      Mohamed, Abdelrahman  and
      Dupoux, Emmanuel  and
      Adi, Yossi",
    booktitle = "Proceedings of the 2022 Conference on Empirical Methods in Natural Language Processing",
    month = dec,
    year = "2022",
    doi = "10.18653/v1/2022.emnlp-main.769",
    pages = "11200--11214"
}

@INPROCEEDINGS{PAD,
  author={Zhou, Kun and Zhang, You and Ng, Dianwen and Zhao, Shengkui and Wang, Hao and Ma, Bin},
  booktitle={ICASSP 2026 - 2026 IEEE International Conference on Acoustics, Speech and Signal Processing (ICASSP)}, 
  title={Emotional Dimension Control in Language Model-Based Text-To-Speech: Spanning a Broad Spectrum of Human Emotions}, 
  year={2026},
  volume={},
  number={},
  pages={17257--17261},
  doi={10.1109/ICASSP55912.2026.11462862}}

@inproceedings{emospheretts,
  title     = {{EmoSphere-TTS: Emotional Style and Intensity Modeling via Spherical Emotion Vector for Controllable Emotional Text-to-Speech}},
  author    = {Deok-Hyeon Cho and Hyung-Seok Oh and Seung-Bin Kim and Sang-Hoon Lee and Seong-Whan Lee},
  year      = {2024},
  booktitle = {{Interspeech 2024}},
  pages     = {1810--1814},
  doi       = {10.21437/Interspeech.2024-398},
  issn      = {2958-1796},
}

@InProceedings{mead,
author="Wang, Kaisiyuan
and Wu, Qianyi
and Song, Linsen
and Yang, Zhuoqian
and Wu, Wayne
and Qian, Chen
and He, Ran
and Qiao, Yu
and Loy, Chen Change",
editor="Vedaldi, Andrea
and Bischof, Horst
and Brox, Thomas
and Frahm, Jan-Michael",
title="MEAD: A Large-Scale Audio-Visual Dataset for Emotional Talking-Face Generation",
booktitle="Computer Vision -- ECCV 2020",
year="2020",
publisher="Springer International Publishing",
address="Cham",
pages="700--717",
isbn="978-3-030-58589-1"
}

@inproceedings{odyssey,
  title     = {{Odyssey 2024 - Speech Emotion Recognition Challenge: Dataset, Baseline Framework, and Results}},
  author    = {Lucas Goncalves and Ali N. Salman and Abinay Reddy Naini and Laureano Moro-Vel{\'a}zquez and Thomas Thebaud and Paola Garcia and Najim Dehak and Berrak Sisman and Carlos Busso},
  year      = {2024},
  booktitle = {{The Speaker and Language Recognition Workshop (Odyssey 2024)}},
  pages     = {247--254},
  doi       = {10.21437/odyssey.2024-35}
}

@article{qwen3,
  title={Qwen3 technical report},
  author={Yang, An and Li, Anfeng and Yang, Baosong and Zhang, Beichen and Hui, Binyuan and Zheng, Bo and Yu, Bowen and Gao, Chang and Huang, Chengen and Lv, Chenxu and others},
  journal={arXiv preprint arXiv:2505.09388},
  year={2025}
}

@article{e5,
  title={Text Embeddings by Weakly-Supervised Contrastive Pre-training},
  author={Wang, Liang and Yang, Nan and Huang, Xiaolong and Jiao, Binxing and Yang, Linjun and Jiang, Daxin and Majumder, Rangan and Wei, Furu},
  journal={arXiv preprint arXiv:2212.03533},
  year={2022}
}

@inproceedings{emo2vec,
    title = "emotion2vec: Self-Supervised Pre-Training for Speech Emotion Representation",
    author = "Ma, Ziyang  and
      Zheng, Zhisheng  and
      Ye, Jiaxin  and
      Li, Jinchao  and
      Gao, Zhifu  and
      Zhang, ShiLiang  and
      Chen, Xie",
    booktitle = "Findings of the Association for Computational Linguistics: ACL 2024",
    month = aug,
    year = "2024",
    pages = "15747--15760",

}

@ARTICLE{hubert,
  author={Hsu, Wei-Ning and Bolte, Benjamin and Tsai, Yao-Hung Hubert and Lakhotia, Kushal and Salakhutdinov, Ruslan and Mohamed, Abdelrahman},
  journal={IEEE/ACM Transactions on Audio, Speech, and Language Processing}, 
  title={HuBERT: Self-Supervised Speech Representation Learning by Masked Prediction of Hidden Units}, 
  year={2021},
  volume={29},
  number={},
  pages={3451--3460},
  doi={10.1109/TASLP.2021.3122291}}

@inproceedings{hifigan,
 author = {Kong, Jungil and Kim, Jaehyeon and Bae, Jaekyoung},
 booktitle = {Advances in Neural Information Processing Systems},
 editor = {H. Larochelle and M. Ranzato and R. Hadsell and M.F. Balcan and H. Lin},
 pages = {17022--17033},
 publisher = {Curran Associates, Inc.},
 title = {HiFi-GAN: Generative Adversarial Networks for Efficient and High Fidelity Speech Synthesis},
 volume = {33},
 year = {2020}
}

@InProceedings{whisper,
  title = 	 {Robust Speech Recognition via Large-Scale Weak Supervision},
  author =       {Radford, Alec and Kim, Jong Wook and Xu, Tao and Brockman, Greg and Mcleavey, Christine and Sutskever, Ilya},
  booktitle = 	 {Proceedings of the 40th International Conference on Machine Learning},
  pages = 	 {28492--28518},
  year = 	 {2023},
  volume = 	 {202},
  series = 	 {Proceedings of Machine Learning Research},
  month = 	 {23--29 Jul},
  publisher =    {PMLR}
}

@inproceedings{utmos,
  title     = {{UTMOS: UTokyo-SaruLab System for VoiceMOS Challenge 2022}},
  author    = {Takaaki Saeki and Detai Xin and Wataru Nakata and Tomoki Koriyama and Shinnosuke Takamichi and Hiroshi Saruwatari},
  year      = {2022},
  booktitle = {{Interspeech 2022}},
  pages     = {4521--4525},
  doi       = {10.21437/Interspeech.2022-439},
  issn      = {2958-1796},
}

@INPROCEEDINGS{prompttts,
  author={Guo, Zhifang and Leng, Yichong and Wu, Yihan and Zhao, Sheng and Tan, Xu},
  booktitle={ICASSP 2023 - 2023 IEEE International Conference on Acoustics, Speech and Signal Processing (ICASSP)}, 
  title={Prompttts: Controllable Text-To-Speech With Text Descriptions}, 
  year={2023},
  volume={},
  number={},
  pages={1--5},
  doi={10.1109/ICASSP49357.2023.10096285}}

@inproceedings{speechcraft,
author = {Jin, Zeyu and Jia, Jia and Wang, Qixin and Li, Kehan and Zhou, Shuoyi and Zhou, Songtao and Qin, Xiaoyu and Wu, Zhiyong},
title = {SpeechCraft: A Fine-Grained Expressive Speech Dataset with Natural Language Description},
year = {2024},
booktitle = {Proceedings of the 32nd ACM International Conference on Multimedia},
pages = {1255--1264},
}

@INPROCEEDINGS{ge2e,
  author={Wan, Li and Wang, Quan and Papir, Alan and Moreno, Ignacio Lopez},
  booktitle={2018 IEEE International Conference on Acoustics, Speech and Signal Processing (ICASSP)}, 
  title={Generalized End-to-End Loss for Speaker Verification}, 
  year={2018},
  volume={},
  number={},
  pages={4879--4883},
  doi={10.1109/ICASSP.2018.8462665}}

@inproceedings{wav2vec,
  title     = {{Unsupervised Cross-Lingual Representation Learning for Speech Recognition}},
  author    = {Alexis Conneau and Alexei Baevski and Ronan Collobert and Abdelrahman Mohamed and Michael Auli},
  year      = {2021},
  booktitle = {{Interspeech 2021}},
  pages     = {2426--2430},
  doi       = {10.21437/Interspeech.2021-329},
  issn      = {2958-1796},
}

@inproceedings{ZSDEVC,
  title     = {{ZSDEVC: Zero-Shot Diffusion-based Emotional Voice Conversion with Disentangled Mechanism}},
  author    = {Hsing-Hang Chou and Yun-Shao Lin and Ching-Chin Sung and Yu Tsao and Chi-Chun Lee},
  year      = {2025},
  booktitle = {{Interspeech 2025}},
  pages     = {4398--4402},
  doi       = {10.21437/Interspeech.2025-1101},
  issn      = {2958-1796},
}

@inproceedings{ClapFM-EVC,
  title     = {{ClapFM-EVC: High-Fidelity and Flexible Emotional Voice Conversion with Dual Control from Natural Language and Speech}},
  author    = {Yu Pan and Yanni Hu and Yuguang Yang and Jixun Yao and Jianhao Ye and Hongbin Zhou and Lei Ma and Jianjun Zhao},
  year      = {2025},
  booktitle = {{Interspeech 2025}},
  pages     = {4583--4587},
  doi       = {10.21437/Interspeech.2025-203},
  issn      = {2958-1796},
}

@inproceedings{spk_indenpend,
  title     = {{Converting Anyone's Emotion: Towards Speaker-Independent Emotional Voice Conversion}},
  author    = {Kun Zhou and Berrak Sisman and Mingyang Zhang and Haizhou Li},
  year      = {2020},
  booktitle = {{Interspeech 2020}},
  pages     = {3416--3420},
  doi       = {10.21437/Interspeech.2020-2014},
  issn      = {2958-1796},
}

@INPROCEEDINGS{textrolspeech,
  author={Ji, Shengpeng and Zuo, Jialong and Fang, Minghui and Jiang, Ziyue and Chen, Feiyang and Duan, Xinyu and Huai, Baoxing and Zhao, Zhou},
  booktitle={ICASSP 2024 - 2024 IEEE International Conference on Acoustics, Speech and Signal Processing (ICASSP)}, 
  title={TextrolSpeech: A Text Style Control Speech Corpus with Codec Language Text-to-Speech Models}, 
  year={2024},
  volume={},
  number={},
  pages={10301--10305},
  doi={10.1109/ICASSP48485.2024.10445879}}

@INPROCEEDINGS{PromptVC,
  author={Yao, Jixun and Yang, Yuguang and Lei, Yi and Ning, Ziqian and Hu, Yanni and Pan, Yu and Yin, Jingjing and Zhou, Hongbin and Lu, Heng and Xie, Lei},
  booktitle={ICASSP 2024 - 2024 IEEE International Conference on Acoustics, Speech and Signal Processing (ICASSP)}, 
  title={Promptvc: Flexible Stylistic Voice Conversion in Latent Space Driven by Natural Language Prompts}, 
  year={2024},
  volume={},
  number={},
  pages={10571--10575},
  doi={10.1109/ICASSP48485.2024.10445804}}

@inproceedings{HybridVC,
  title     = {{HybridVC: Efficient Voice Style Conversion with Text and Audio Prompts}},
  author    = {Xinlei Niu and Jing Zhang and Charles Patrick Martin},
  year      = {2024},
  booktitle = {{Interspeech 2024}},
  pages     = {4368--4372},
  doi       = {10.21437/Interspeech.2024-46},
  issn      = {2958-1796},
}

@inproceedings{nisqa,
  title={{NISQA: A Deep CNN-Self-Attention Model for Multidimensional Speech Quality Prediction with Crowdsourced Datasets}},
  author={Mittag, Gabriel and Naderi, Babak and Chehadi, Assmaa and M{\"o}ller, Sebastian},
  booktitle={Interspeech 2021},
  pages={2127--2131},
  year={2021}
}

@inproceedings{libritts,
  title     = {{LibriTTS: A Corpus Derived from LibriSpeech for Text-to-Speech}},
  author    = {Heiga Zen and Viet Dang and Rob Clark and Yu Zhang and Ron J. Weiss and Ye Jia and Zhifeng Chen and Yonghui Wu},
  year      = {2019},
  booktitle = {{Interspeech 2019}},
  pages     = {1526--1530},
  doi       = {10.21437/Interspeech.2019-2441},
  issn      = {2958-1796},
}

@article{hdd,
  title={MTLD, vocd-D, and HD-D: A validation study of sophisticated approaches to lexical diversity assessment},
  author={McCarthy, Philip M and Jarvis, Scott},
  journal={Behavior research methods},
  volume={42},
  number={2},
  pages={381--392},
  year={2010},
  publisher={Springer}
}

@inproceedings{zhu2018texygen,
  title={Texygen: A benchmarking platform for text generation models},
  author={Zhu, Yaoming and Lu, Sidi and Zheng, Lei and Guo, Jiaxian and Zhang, Weinan and Wang, Jun and Yu, Yong},
  booktitle={The 41st international ACM SIGIR conference on research \& development in information retrieval},
  pages={1097--1100},
  year={2018}
}

@article{minicpm,
  title={Minicpm-o 4.5: Towards real-time full-duplex omni-modal interaction},
  author={Cui, Junbo and Xu, Bokai and Wang, Chongyi and Yu, Tianyu and Sun, Weiyue and Xu, Yingjing and Wang, Tianran and He, Zhihui and Ma, Wenshuo and Cai, Tianchi and others},
  journal={arXiv preprint arXiv:2604.27393},
  year={2026}
}

@article{qwen3-omni,
  title={Qwen3-omni technical report},
  author={Xu, Jin and Guo, Zhifang and Hu, Hangrui and Chu, Yunfei and Wang, Xiong and He, Jinzheng and Wang, Yuxuan and Shi, Xian and He, Ting and Zhu, Xinfa and others},
  journal={arXiv preprint arXiv:2509.17765},
  year={2025}
}

@article{kimi,
  title={Kimi-audio technical report},
  author={Ding, Ding and Ju, Zeqian and Leng, Yichong and Liu, Songxiang and Liu, Tong and Shang, Zeyu and Shen, Kai and Song, Wei and Tan, Xu and Tang, Heyi and others},
  journal={arXiv preprint arXiv:2504.18425},
  year={2025}
}

@article{MERaLiON-SER,
  title={MERaLiON-SER: Robust Speech Emotion Recognition Model for English and SEA Languages},
  author={MERaLiON Team},  journal={http://arxiv.org/abs/2511.04914},
  year={2025}
}

@inproceedings{voxinstruct,
  title={Voxinstruct: Expressive human instruction-to-speech generation with unified multilingual codec language modelling},
  author={Zhou, Yixuan and Qin, Xiaoyu and Jin, Zeyu and Zhou, Shuoyi and Lei, Shun and Zhou, Songtao and Wu, Zhiyong and Jia, Jia},
  booktitle={Proceedings of the 32nd ACM International Conference on Multimedia},
  pages={554--563},
  year={2024}
}

@inproceedings{instructpix2pix,
  title={Instructpix2pix: Learning to follow image editing instructions},
  author={Brooks, Tim and Holynski, Aleksander and Efros, Alexei A},
  booktitle={2023 IEEE/CVF Conference on Computer Vision and Pattern Recognition (CVPR)},
  pages={18392--18402},
  year={2023},
  organization={IEEE}
}

@article{tau,
  title   = {A New Measure of Rank Correlation},
  author  = {Kendall, M. G.},
  journal = {Biometrika},
  volume  = {30},
  number  = {1-2},
  pages   = {81--93},
  year    = {1938},
  doi     = {10.1093/biomet/30.1-2.81}
}

@inproceedings{emorank,
  title     = {{EmotionRankCLAP: Bridging Natural Language Speaking Styles and Ordinal Speech Emotion via Rank-N-Contrast}},
  author    = {Shreeram Suresh Chandra and Lucas Goncalves and Junchen Lu and Carlos Busso and Berrak Sisman},
  year      = {2025},
  booktitle = {{Interspeech 2025}},
  pages     = {3000--3004},
  doi       = {10.21437/Interspeech.2025-1198},
  issn      = {2958-1796},
}

\end{document}